# Are University Rankings Statistically Significant?
# A Comparison among Chinese Universities and with the USA

Loet Leydesdorff[1], Caroline S. Wagner[2], and Lin Zhang* [3]

[1] Amsterdam School of Communication Research (ASCoR), University of Amsterdam, PB 15793, 1001 NG Amsterdam, The Netherlands; l.a.leydesdorff@uva.nl;

[2] John Glenn College of Public Affairs, The Ohio State University, Columbus, Ohio, USA, 43210; wagner.911@osu.edu

[3] School of Information Management, Wuhan University, Wuhan, China; linzhang1117@whu.edu.cn

**Abstract**

**Purpose:** We address the question of whether differences are statistically significant in the rankings of universities. We propose methods measuring the statistical significance among different universities and illustrate the results by empirical data.

**Design/methodology/approach:** Based on *z*-testing and overlapping confidence intervals, and using data about 205 Chinese universities included in the Leiden Rankings 2020, we argue that three main groups of Chinese research universities can be distinguished.

**Findings:** When the sample of 205 Chinese universities is merged with the 197 US universities included in Leiden Rankings 2020, the results similarly indicate three main groups: high, middle, low. Using this data (Leiden Rankings and Web-of-Science), the *z*-scores of the Chinese universities are significantly below those of the US universities albeit with some overlap.

**Research limitations:** We show empirically that differences in ranking may be due to changes in the data, the models, or the modeling effects on the data. The scientometric groupings are not always stable when we use different methods.

**R&D policy implications:** Differences among universities can be tested for their statistical significance. The statistics relativize the values of decimals in the rankings. One can operate with a scheme of low/middle/high in policy debates and leave the more fine-grained rankings of individual universities to operational management and local settings.

**Originality/value:** In the discussion about the rankings of universities, the question of whether differences are statistically significant, is, in our opinion, insufficiently addressed.

* Corresponding author: Lin Zhang, linzhang1117@whu.edu.cn.

## 1. Introduction

Classifications and rankings are based on assumptions and decisions about parameters. For example, Harvard University is listed at the top in many rankings. However, when one controls for the budget by dividing the numbers of publications and/or citations (output) by budget (input), other universities come to the fore. Leydesdorff & Wagner (2009), for example, found Eastern-European universities (Poland, Slovakia) as most efficient in terms of output/dollar, because of the relatively low costs of skilled labour in these countries at the time. The ranked order is conditioned by the choice of indicators.

In China, for example, the mathematics department of Qufu Normal University unexpectedly led the ranking of US News with a 19th position worldwide.[1] This unexpected results generated a heated discussion about rankings. Qufu Normal University is a provincial university. How could it be ahead of Peking University, which is traditionally believed to have the strongest mathematics department in China. Shandong University of Science and Technology was ranked third, trailing Peking University by only 0.1 points. (Tsinghua University ranked sixth.) However, the majors in mathematics of these two universities are rated relatively low when using other indicators including international collaborations. In sum, a rank-order is conditioned by the choice of indicators and by how the various indicators are weighed in the attribution (Shin, Toutkoushian, & Teichler, 2011).

Comparing universities may be even more complex than comparing nations. While nations can be expected to cover portfolios of standardized measurements,[2] universities can be specialized by discipline or by mission (e.g., agricultural universities). What counts as 'output' of universities is debatable. Furthermore, universities provide higher-education, and one can argue that this task should be appreciated in a ranking, even if it is difficult to measure. University ranking services seek to compare such heterogeneous indicators by weighting, summing, or averaging partial indicators, but often without giving much value to mission-orientation and social relevance.

The reliability and reproducibility of yearly rankings is low because the models are based on specific choices in a parameter space of possible choices. Gingras (2016, at p. 75), for example, concluded "that annual rankings of universities, be they based on surveys, bibliometrics, or webometrics, have no foundation in methodology and can only be explained as marketing strategies on the part of the producers of these rankings." Others have pointed to the reputational effects in rankings. Rankings can instantiate historical power structures (Bowman & Bastedo, 2017). However, numbers can also provide a new perspective because ongoing changes may have gone hitherto unnoticed. We shall provide examples of such unexpected developments below.

[1] https://www.usnews.com/education/best-global-universities/china/mathematics; retrieved 27 October 2020.
[2] See, for example, OECD's Frascati Manual: https://www.oecd.org/publications/frascati-manual-2015-9789264239012-en.htm

In our opinion, statisticians and bibliometricians have made progress during the last decade in developing arguments for making informed decisions about the data and relevant methodologies for comparing institutions. The rankings no longer have to be arbitrary and subjective combinations of surveys and partial indicators. However, they remain constructs which enable us to distinguish among universities in specific terms. For example, the Leiden Rankings (LR) of research universities stand out in terms of their transparency and clear objectives. The objective of the LR exercises is to rank research universities exclusively in terms of their publications and citations as output, without accounting for other differences.

## 2. Decisions about data

In addition to providing the rankings online (at https://www.leidenranking.com/ranking/2020/list), the yearly source data for LR are freely available in an Excel file for secondary analysis. This data is derived from the Web-of-Science of the Institute of Scientific Information (ISI/ Clarivate) in Philadelphia. The address information is disambiguated and reorganized by the Centre for Science and Technology Studies (CWTS) in Leiden.

LR does not take into account conference proceedings and book publications; it is exclusively based on research articles and reviews published in the journals included in the Web-of-Science.[3] Consequently, the coverage may favor work in the natural and engineering sciences more than the social science and humanities (Sivertsen, 2016). However, this data can be considered as an attempt to capture the most intellectual contribution to knowledge production among the numerous outputs of research universities (Garfield, 1971). A major decision remains thereupon to attribute coauthored papers for a full count to each of the (co-)authors and their home institutions or to attribute this credit proportionally (so-called “fractional counting”); both of these methods are made available in the Excel files.

[3] Clarivate's criterion for classifying papers as reviews is as follows: "In the *JCR* system any article containing more than 100 references is coded as a review. Articles in 'review' sections of research or clinical journals are also coded as reviews, as are articles whose titles contain the word 'review' or 'overview'" (at http://thomsonreuters.com/products_services/science/free/essays/impact_factor/ (retrieved 8 April 2012; https://clarivate.com/webofsciencegroup/essays/impact-factor/ (retrieved on 29 August 2020).

The full counting method gives a full point to each author and her institutional address. Fractional counting divides the attribution by the number of authors or institutions. Bibliometrically, fractional counting has the advantage that each publication is eventually counted as one full point and thus percentages add up to hundred (Anderson, 1988). Among the possible counting rules, LR uses fractional counting at the institutional level. For example, if a publication is co-authored by three researchers and two of these researchers are affiliated with a particular university, the publication has a weight of 2 / 3 = 0.67 in the attribution of the scientific impact to this university (Sivertsen, Rousseau, & Zhang, 2019; p. 680).

Another important decision is the time window used to harvest publications and citations, respectively. LR uses four-year periods for cited publications. The last interval for LR 2020 is 2015-2018; the count included citations to publications in these four years cumulatively. The resulting file with input data for this study contained ranks for 1,176 research universities in 65 countries in the preceding years in intervals of four years (Table 1). Rankings are provided both fractionally and as whole counts. In terms of disciplines, data is provided for “All sciences” and five major fields on the basis of relevant journal categories: (*i*) biomedical and health sciences, (*ii*) life and earth sciences, (*iii*) mathematics and computer science, (*iv*) physical sciences and engineering, and (*v*) social sciences and humanities.

**Table 1**: Number of research universities included in LR 2020 for respective countries.

| **Country (top-10 of 65)** | ***N* of universities** | |
|---|---|---|
| China | 205 | |
| United States | 198 | |
| United Kingdom | 58 | |
| Germany | 54 | |
| Japan | 53 | |
| South Korea | 44 | |
| Italy | 41 | |
| Spain | 41 | |
| India | 36 | |
| Iran | 36 | % |
| sum | 766 | 65.1% |
| <55 other countries> | = (1176 – 766) = 410 | 34.9% |
| 65 countries | 1176 | 100% |

In this study, we limit the analysis first to "All sciences," the last available period (2015-2018), and fractional counting. However, this analysis can be repeated analogously using subsets with other parameter choices. Our sample contains 205 Chinese universities (see Table 2) covered by LR 2020. This data was further processed by us with dedicated routines, which are online available at http://www.leydesdorff.net/software/leiden (Leydesdorff *et al*., 2019). These routines analyze the comparisons in each nation. China and the US, however, are represented in LR 2020 with 205 and 198 universities, respectively, making comparisons possible. For example, one can compare Harvard with Stanford as US universities, and—as we shall see below—it is possible to compare them with Tsinghua or Zhejiang University.

## 3. Methods

### *3.1. Statistical Significance*

The values of indicators are almost by definition unequal between two measurements, but one can ask whether differences are statistically significant or fall within the margin of error. In a study about German, British, and US universities, Leydesdorff, Bornmann, & Mingers (2019) proposed three statistics for distinguishing among groups of universities: (*i*) overlapping confidence intervals, (*ii*) *z*-tests based on differences between observed and expected values of the percentage publications in the top-10% group (PP-top10%)—and (*iii*) effect sizes (Cohen, 1988). Although many statisticians nowadays have a preference for the latter measure (Schneider, 2015; Waltman, 2016), effect sizes are less known to practitioners. Power analysis based on effect sizes is sometimes considered to be of "practical significance" (Cumming, 2013; Wasserstein & Lazar, 2016). However, the scientometric interpretations of the effect sizes in our previous study were not convincing.

We limit the discussion here to *z*-testing and overlapping confidence intervals. When the differences between two universities are not significantly different, they can be grouped together. Using the *z*-scores or the relative overlaps, one can test the differences and thus generate groups of nodes with links among them indicating group membership. These results can be analyzed and visualized as clusters. The *z*-scores can also be used as quantitative measures of differences among nodes (universities) and links (between universities).

### *3.2. Observed versus Expected*

Most commonly, one tests for the significance of the differences between *mean* values of variables or, in the case of citation analysis, between the so-called c/p-ratios—that is, the *mean* numbers of citations per publication. However, scientometric distributions are highly skewed; the mean is therefore not a meaningful indicator of the central tendency in the distribution.[4] An alternative could be found in using the median which is by definition equal to the top-50% of the distribution. Given the skew of the distribution, however, quality indicators can also focus on the numbers of top-10% or even top-1% most-highly cited papers (Bornmann & Mutz, 2011; Leydesdorff, Bornmann, Mutz, & Opthof, 2011; cf. McAllister, Narin, & Corrigan, 1983; Tijssen, Visser, & Van Leeuwen, 2002).

Counts allow for significance testing of the differences between expected and observed values using non-parametric statistics. For example, chi-square is formulated as follows:

$$\chi^2 = \sum_{i=1}^{n} \frac{(Observed_i - Expected_i)^2}{Expected_i} \quad (1)$$

Significance of the resulting chi-square values can be looked-up in any table of chi-square values; for example, at the Internet.

Additionally, and in this context importantly, the individual terms before the squaring $[\frac{(Observed_i - Expected_i)}{\sqrt{Expected_i}}]$ are the so-called standardized residuals of the chi-square. Any residual with an absolute value > 1.96 is significant at the 5% level, and any residual > 2.576 is significant at the 1% level. In other words, the residuals are *z*-scores enabling a detailed investigation of differences at the level of each individual cell of a matrix.

Let us elaborate with a numerical example comparing Tsinghua and Zhejiang University. Table 2a shows the values obtained from LR 2020 data for these two leading universities.

**Table 2a**: Observed values of top-10% cited papers for Tsinghua and Zhejiang University during the period 2015-2019 (fractional counting, all sciences)

| *Observed values* | *top-10%* | *non-top* | *total p* |
|---|---|---|---|
| *Tsinghua* | 2738 | 17164 | 19902 |
| *Zhejiang* | 2604 | 20906 | 23510 |
| | 5342 | 38070 | 43412 |

[4] The so-called "crown indicator" in scientometrics (van Raan *et al* , 2010; Waltman *et al*., 2011) is unfortunately the *Mean Normalized Citation Score* (MNCS). As noted, the *mean* is an unfortunate choice as a central-tendency statistic of a skewed distribution.

The expected values can be derived from the observed ones by using the margin-totals and grand-total of the cross-table as follows: *Expectation*(ij) = (Σ.j Σi.) / Σ..). In other words, the product of the column total and the row total is divided by the grand total of the matrix. Applying this counting rule, the expected number of papers in the top-10% category of Tsinghua University is (19902 * 5342) / 43412 = 2449.01. This value is written in the top-left cell of Table 2b.

**Table 2b**: Expected values of top-10% cited papers for Tsinghua and Zhejiang University

| *Expected values* | *top-10%* | *non-top* | *total p* |
|---|---|---|---|
| *Tsinghua* | 2449.01 | 17452.99 | 19902.00 |
| *Zhejiang* | 2892.99 | 20617.01 | 23510.00 |
| | 5356.09 | 38055.91 | 43412.00 |

Using Eq. 1, Table 3 shows the contributions of cell values to the chi-square. The sum of the values in Table 3 is the chi-square; in this case 71.80. The corresponding *p*-value is < 0.001 and thus the differences are statistically significant.

**Table 3:** Chi-square for the comparison of Tsinghua and Zhejiang University in LR 2020

| *Chi-square* | *top-10%* | *non-top* | |
|---|---|---|---|
| *Tsinghua* | 34.10 | 4.79 | 38.89 |
| *Zhejiang* | 28.87 | 4.05 | 32.92 |
| | | $\chi^2$ = | **71.80** |

**Table 4:** Standardized residuals of the chi-square values in Table 4

| | Top-10% | Non-top |
|---|---|---|
| *Tsinghua* | 5.84 | -2.19 |
| *Zhejiang* | -5.37 | 2.01 |

Table 4 adds the residuals of the chi-square for this data. Tsinghua ranks significantly above expectation in the top-10% category (z > 2.576), and non-significantly below expectation in the other publications. For Zhejiang the opposite is the case. In conclusion: these two universities cannot be considered statistically as belonging to the same group.

### *3.3. z-test*

Without prior knowledge of historical or social contexts, one would expect that 10% of a university's publications will belong to the 10% most-highly cited papers in the reference group. A university that publishes more than 10% of these top papers scores above expectation. The z-test can be used to test the significance of the observed number of papers in the top-10% segment against the expected 10%. In addition to comparing a university with the expectation, the test can be applied to the differences between any two universities. A value of $z$=1.96 indicates a significance of the difference at the 5% level: $z > 2.576$ indicates that the chance process is only 1 in 100, and for $z > 3.29$, the chance rate is only one *per mille*. A negative z-score indicates *mutatis mutandis* that the score is *below* expectation. The resulting $z$ values can be compared and used for ranking purposes.

It can be derived from Eq.1 that the test statistics between two proportions—percentages are proportions—can be formulated as follows (Sheskin, 201, pp. 656f.):

$$z = \frac{p_1 - p_2}{\sqrt{p(1-p)\left[\frac{1}{n_1} + \frac{1}{n_2}\right]}} \qquad (2)$$

where $n_1$ and $n_2$ are the numbers of all the papers published by institutions 1 and 2 (under the column "$P$" in the LR); and $p_1$ and $p_2$ are the values of $PP_{top\ 10\%}$ of institutions 1 and 2. The pooled estimate for proportion $p$ is defined as:

$$p = \frac{t_1 + t_2}{n_1 + n_2} \qquad (3)$$

where: $t_1$ and $t_2$ are the numbers of top-10% papers of institutions 1 and 2. These numbers can be calculated on the basis of the values for "$P$" and "$PP_{top\ 10\%}$" in LR. When testing values for a single university, $n_1 = n_2$, $p_1$ is the value of the $PP_{top\ 10\%}$, $p_2 = 0.1$, and $t_2 = 0.1 * n_2$ (that is, the expected number in the top-10%).

Using the same numerical example as above, $n_1$ = 19,902 for Tsinghua and $n_2$ = 23,510 for Zhejiang, respectively; $t_1$ = 2738 and $t_2$ = 2604 (Table 3a above). The pooled estimate $p$ is in this case:

$$p = \frac{2738+2604}{19902+23510} = \frac{5342}{43412} = 0.123054$$

$$p\ (1 - p) = 0.1234 * 0.8766 = 0.1081$$

Using Eq. 2, one can fill out as follows:

$$z = \frac{\frac{13.8-11.1}{100}}{\sqrt{0.1081\left[\frac{1}{19,902} + \frac{1}{23,510}\right]}} = 8.525$$

The *z*-test indicates that the scores of these two universities are statistically different above the 0.001 level.

It can be shown that if both the z-test and the chi-square are applied to the same set of data, the square of the *z*-value is equal to the chi-square value (Sheshkin, 2011, p. 655). The residuals to the chi-square are standardized as a *z*-statistics as well. The *z*-test for two independent proportions (e.g., percentages) provides an alternative large-samples procedure for evaluating contingency tables. The *z*-test is the most appropriate test given the research questions and the design (Sheshkin, 2011, pp. 671f.). [5]

### *3.4. Confidence intervals*

The LR additionally provides confidence intervals[6] which can be used as another statistic for grouping universities.[7] When the confidence intervals of two universities overlap, the distinction between them can be ignored in terms of the indicator (e.g., Colliander & Ahlgren, 2011, at p. 105). The words confidence and stability intervals can be used interchangeably.

Since each of two universities may be indistinguishable from other universities, one thus obtains a so-called "weak" component in terms of network analysis. If both the upper and lower bounds of university A are contained within the stability interval of university B, the performance of the former can be seen as similar to the latter; network analysis would place them into the same cluster. In this case, we have a strong component since both arcs are valued.

Using the same example of comparing Tsinghua with Zhejiang, Figure 1 shows that there is no overlap between their confidence intervals. This accords with our previous conclusion that the output of these two universities in terms of *PP*-top10% is significantly different. For didactic reasons, we added Peking University to the comparison in Figure 1 and Table 7 so that we can draw Figure 2 with the *z*-values as a further illustration of the options for visualization.

---

[5] At http://www.leydesdorff.net/leiden11/index.htm the user can retrieve a file leiden11.xls which allows for feeding values harvested from the LR for the comparison of any two universities. The effect sizes are additionally provided in the template.

[6] The confidence intervals are based on bootstrapping; that is, random drawings that are sufficiently repeated to provide stable patterns. In case of the LR, one draws thousand times a sample from each university's set of publications. In order to obtain a 95% stability interval, the lower and upper bounds of the stability intervals are taken as the 2.5th and the 97.5th percentiles of the thus generated distribution of $PP_{top\ 10\%}$ values (Waltman *et al.*, 2012, at p. 2429).

[7] Bornmann *et al.*'s (2013) analysis of LR 2011 compared the stability intervals with other possible ways to calculate standard errors (e.g., the standard errors of a binary probability). They found a perfect correlation between stability intervals and these other possible ways which are based on less data-intensive computing procedures.

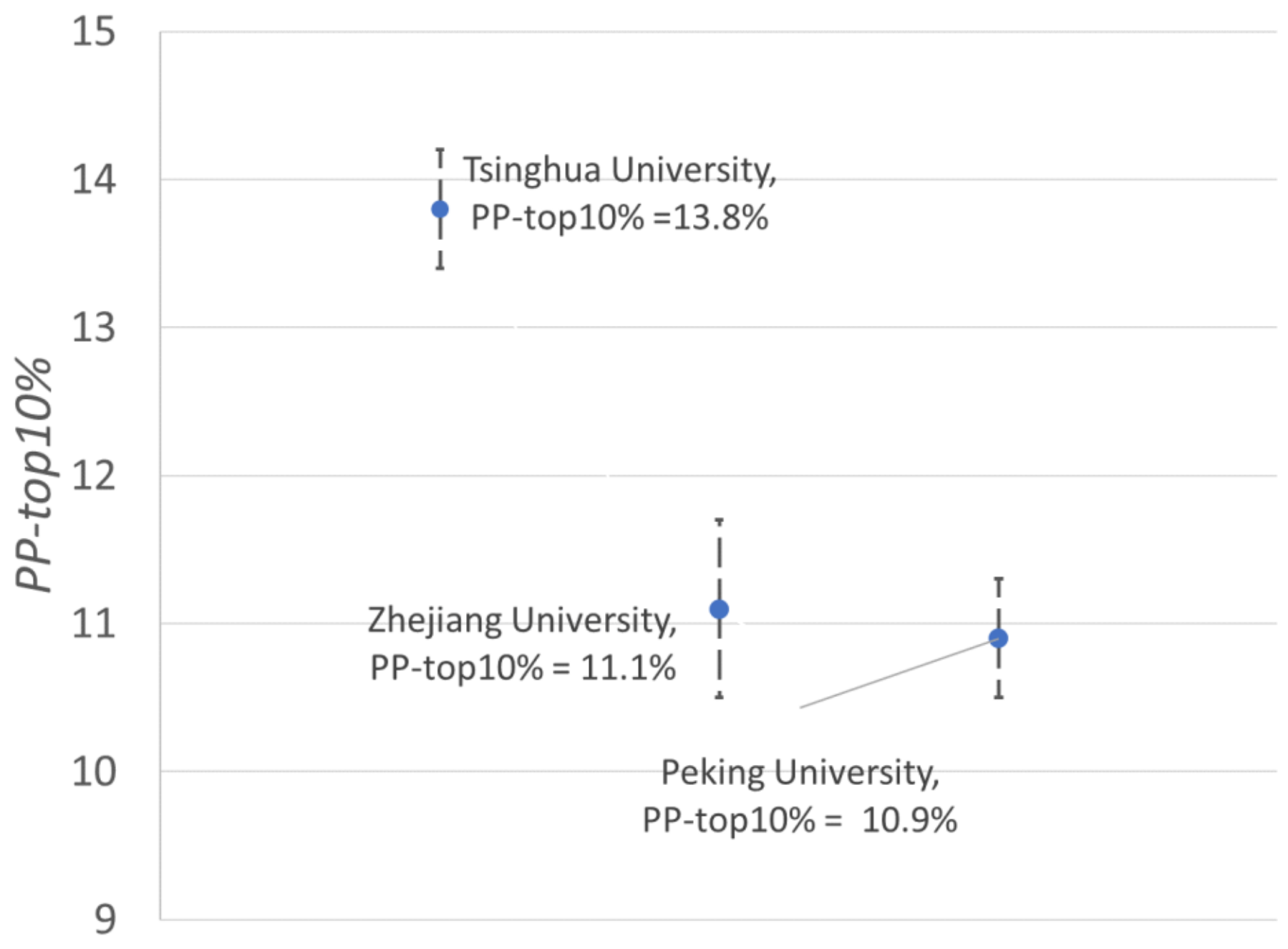

**Figure 1**: Potentially overlapping confidence intervals of the PP-Top10% for three leading Chinese universities.

Zhejiang and Peking University both have a lower bound of 10.5% of top 10% articles. The upper bound of Zhejiang, however, is higher than for Peking University. Table 5 shows the *z*-values for these three nodes on the main diagonal and the links off-diagonal. The *z*-values are also written into Figure 2.

**Table 5**: *z*-values for nodes and links among three leading Chinese universities

| | Peking University | Tsinghua University | Zhejiang University |
|---|---|---|---|
| Peking University | 2.689 ** | | |
| Tsinghua Univ. | 8.460 *** | 11.005 *** | |
| Zhejiang University | 0.638 | 8.533 *** | 3.800 *** |

Significance levels: * $p < .05$; ** $p < .01$; *** $p < .001$

The difference between Peking and Zhejiang University is *not significant*, leading to an arc between these two universities ($z = 0.64$) in Figure 2 and placing them into a cluster together. This weak component does not include Tsinghua University which scores significantly different on this performance indicator (PP-Top10%).

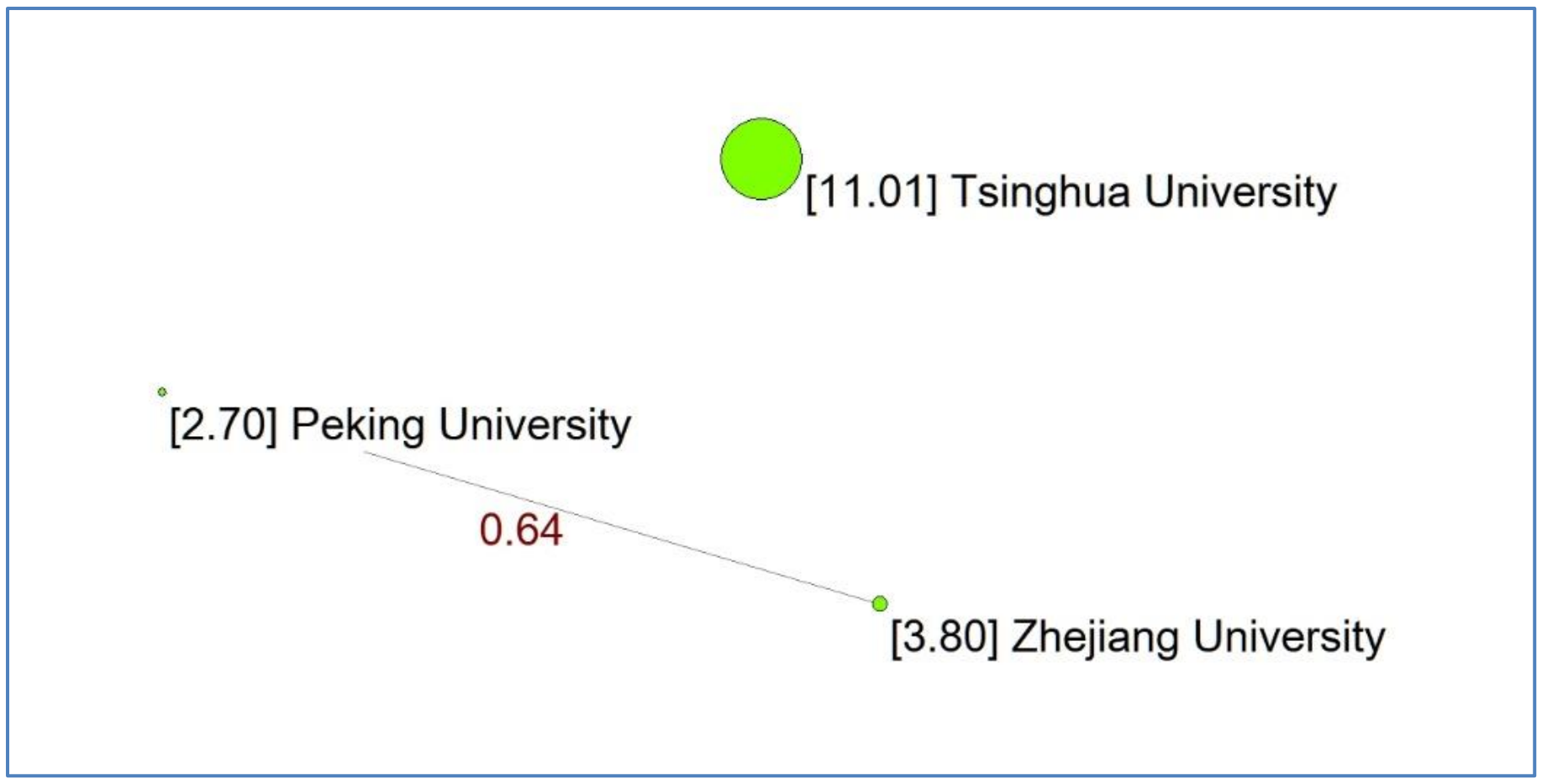

**Figure 2**: Grouping of significantly different and non-different values for *PP*-top10% among three Chinese universities; *z*-values (provided in the figure) are used for sizing the nodes.

# 4. Results

### *4.1. Results based on using the z-test*

Aggregating universities into networked groups based upon their z-score similarities creates clusters. Figure 3 shows the resulting clusters among 205 Chinese universities. The *z*-values are used as input to the sizes of the nodes and fonts, and the *z*-values between two universities determine the lines insofar as $z < 2.576$ ($p<.0.01$) since universities which are not significantly different, are considered as part of the same group. Note that this grouping is on the basis of (structural) similarity and not on actions. We use VOSviewer only for the visualization, but the grouping is based on the above statistics. The file is analytically organized external to VOSviewer.

**Figure 3**: Grouping of 205 Chinese universities in terms of *z*-values ($p < .01$);
VOSviewer used for the decomposition and clustering; modularity Q = 0.165.
The figure can be web-started from [here]() (or with a black background from [here]()).

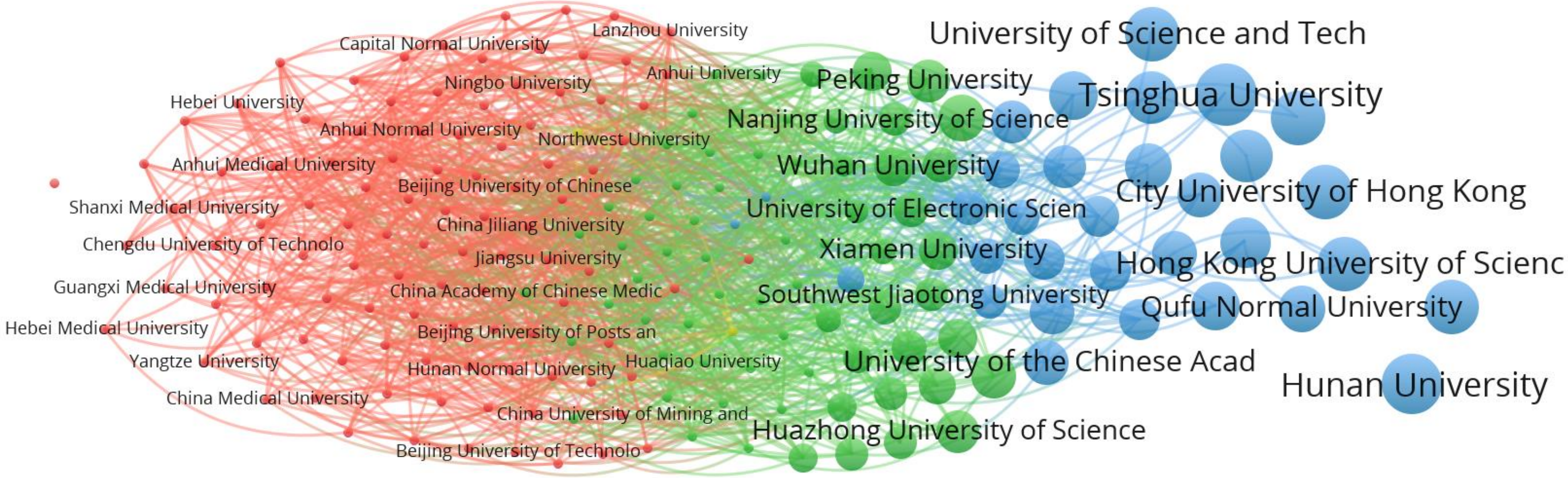

Three major groups of universities and two isolates are distinguished in the analysis and shown in Figure 3: A top-group of 32 universities is listed in Annex 1 (in the supplementary materials of this paper). As could be expected, Tsinghua leads this group with a *z*-score of 11.0, followed by Hunan University ($z$ = 10.2).

The second group of 69 universities is headed by Zhejiang University and listed in Annex 2. As shown above, Peking University is not significantly different from Zhejiang University; it is ranked on the 24th position overall but, following Zhejiang, Peking University is at the 5th position within the second group. Only the top-30 (43.5%) of these 69 universities score on the *PP*-10% above expectation. A third group of 102 universities are listed in Annex 3. None of these universities score above expectation.

Two universities—Chang'an and Shanxi—form a fourth cluster with both negative *z*-values. Table 6 provides the top-20 universities in each of the three groups in decreasing order (on the basis of the *z*-values). Note that five of the 20 universities listed on the top-list are from an address in Hong Kong.

**Table 6:** Top-20 universities in each of the three clusters. (See for a full list in the Annexes.)

| | *top group (top 20)* | *z* | *middle group (top 20)* | *z* | *bottom group (top 20)* | *z* |
|---|---|---|---|---|---|---|
| 1 | Tsinghua Univ. | 11.005 | Zhejiang Univ. | 3.800 | Hubei Univ. | -0.498 |
| 2 | Hunan Univ. | 10.193 | Harbin Institute of Technology | 3.226 | Northwest Univ. | -1.609 |
| 3 | Hong Kong Univ. of Science and Technology | 6.566 | Huazhong Univ. of Science and Technology | 2.907 | South China Agricultural Univ. | -1.631 |
| 4 | Univ. of Science and Technology of China | 6.482 | Peking Univ. | 2.689 | Zhejiang Univ. of Technology | -1.676 |
| 5 | City Univ. of Hong Kong | 6.454 | Nanjing Univ. | 1.973 | China Univ. of Mining and Technology | -1.678 |
| 6 | Shandong Univ. of Science and Technology | 6.444 | Xiamen Univ. | 1.954 | Nanjing Univ. of Aeronautics and Astronautics | -1.816 |
| 7 | Hong Kong Polytechnic Univ. | 6.406 | Wuhan Univ. | 1.765 | Second Military Medical Univ. | -1.844 |
| 8 | South China Univ. of Technology | 6.049 | Tianjin Univ. | 1.738 | Zhejiang Sci-Tech Univ. | -1.886 |
| 9 | Chinese Univ. of Hong Kong | 5.993 | Dalian Univ. of Technology | 1.553 | Taiyuan Univ. of Technology | -1.934 |
| 10 | Nankai Univ. | 4.860 | East China Normal Univ. | 1.333 | Shanghai Univ. of Traditional Chinese Medicine | -1.950 |
| 11 | Univ. of Hong Kong | 4.418 | Soochow Univ. | 1.310 | Qingdao Univ. of Science and Technology | -1.961 |
| 12 | Shenzhen Univ. | 4.089 | Univ. of Electronic Science and Technology of China | 1.154 | Nanjing Univ. of Chinese Medicine | -1.981 |
| 13 | Qufu Normal Univ. | 3.724 | Nanjing Univ. of Science and Technology | 1.140 | Harbin Engineering Univ. | -2.033 |
| 14 | Fuzhou Univ. | 3.487 | Southwest Jiaotong Univ. | 1.088 | Lanzhou Univ. | -2.110 |
| 15 | Beihang Univ. | 3.432 | China Univ. of Geosciences | 1.035 | Northeast Normal Univ. | -2.128 |
| 16 | Univ. of the Chinese Academy of Sciences | 3.354 | Huazhong Agricultural Univ. | 1.030 | China Jiliang Univ. | -2.130 |
| 17 | Central China Normal Univ. | 3.090 | Tongji Univ. | 0.954 | Jiangnan Univ. | -2.163 |
| 18 | Univ. of Macau | 3.046 | Univ. of Science and Technology Beijing | 0.932 | Jinan Univ. | -2.227 |
| 19 | Northwestern Polytechnical Univ. | 2.915 | China Agricultural Univ. | 0.704 | Hangzhou Normal Univ. | -2.245 |
| 20 | Wuhan Univ. of Technology | 2.914 | Beijing Institute of Technology | 0.587 | Southern Medical Univ. | -2.263 |

The results are sometimes counterintuitive. However, Brewer *et al.* (2001) drew attention to the difference between prestige and reputation; prestige is sticky, whereas the citation windows are only four years in LR. For example, Fudan University is considered a prestigious university in China. In the period under study, however, Fudan was listed as an address in 15,442 papers in journals included in the ISI-list. Only 1,395 of these papers (or 9.0%) belonged to the top-10% most-cited papers. This profile is not significantly different from other universities in the *third* group.

The reason for this relatively low rank for Fudan University is not a decline of publications among the addresses, but a relative decline of papers in the top-10% in this university's publications. The number of publications with Fudan University among the author-addresses shows an increase of 1,127 papers over the consecutive four-year periods and on the basis of fractional counting (Figure 4). However, the yearly increase in the number of publications in the top-10% segment is only 6.9% (775 papers/ year). The relative decline can be cumulative as a composed interest rate (Fig. 4). It may be caused by all kind of effects in the data or in the model. *Ceteris paribus*, for example, an increase in fractional counting leads to a decline in the share of publications and citations.

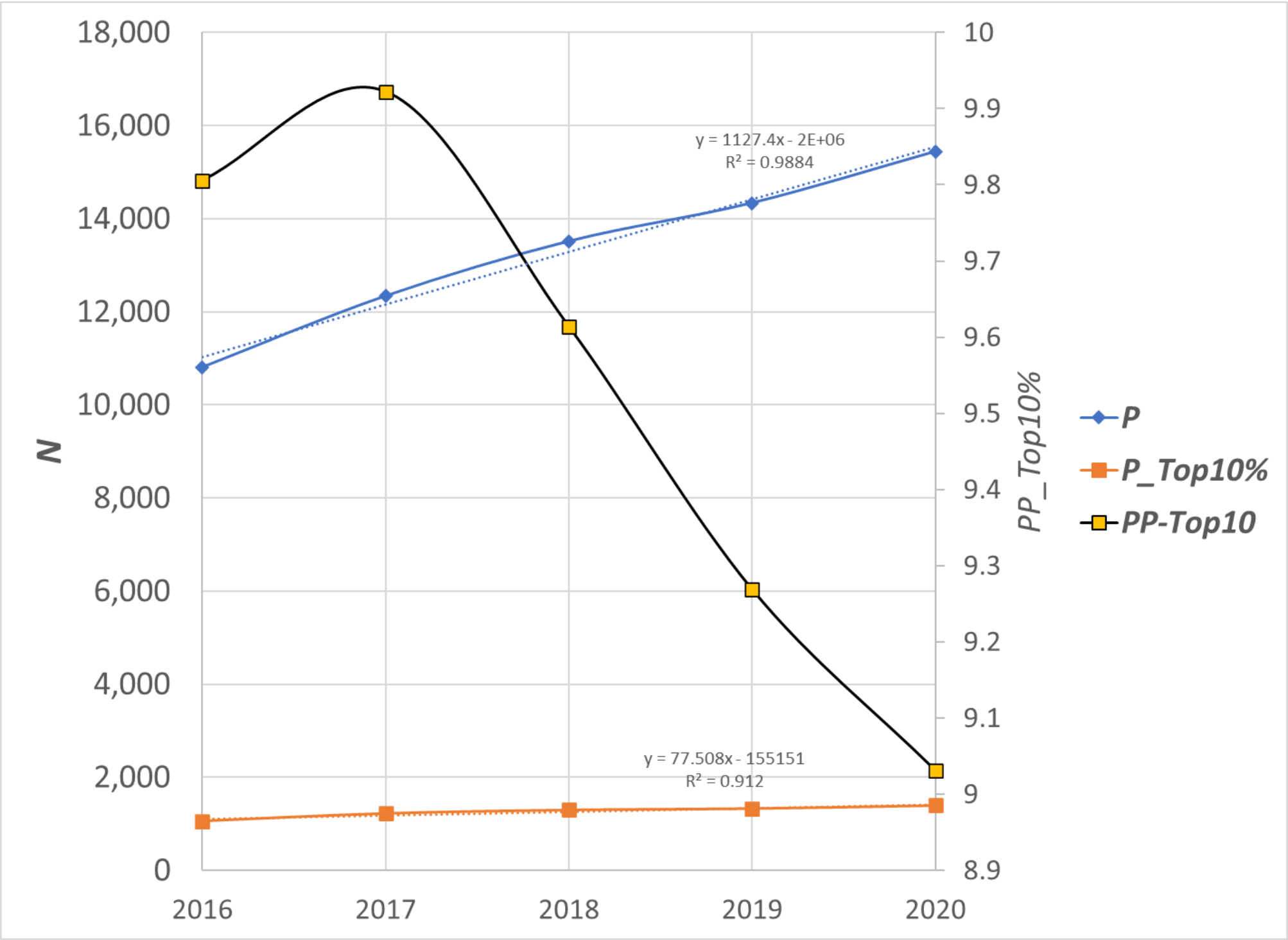

**Figure 4:** Development of the values of *P, P-top10%, and PP-Top10%* for Fudan University during 2016-2020.

*Mutatis mutandis*, one can be puzzled by the high status of Shandong University of Science and Technology at the sixth position and Qufu Normal University at the 13th on the top-list. Shandong University published 1,576 articles (using fractional counting) during the period under study, of which 299 belong to the top-10%. This is almost 19% and thus far exceeds the expectation of 10%. Further analysis may enable us to understand these counterintuitive results.

### *4.2. Results based on confidence intervals*

Figure 5 provides the resulting figure using the overlaps in confidence levels for the delineation of groups: 75 universities are classified as top-universities. These universities are listed in Annex 4.

A measure for the correspondence between the two classifications—the one above based on *z*-scores (Figure 2) and this one (Figure 5 below)—is provided by Cramèr's *V*, which is based on chi-square statistics, but which conveniently varies between zero and one. Cramèr's *V* between these two classifications is significant ($V = 0.48$; $p < 0.01$).[8]

[8] An alternative measure is phi; phi = 0.831 ($p < 0.01$). The Spearman rank-order correlation between the clustering based on the two methods is .6 ($p < 0.01$).

**Figure 5:** Grouping of 205 Chinese universities based on overlapping confidence intervals; VOSviewer used for the decomposition and clustering; font- and node-sizes based on *z*-values. (The map can be web-started from here.)

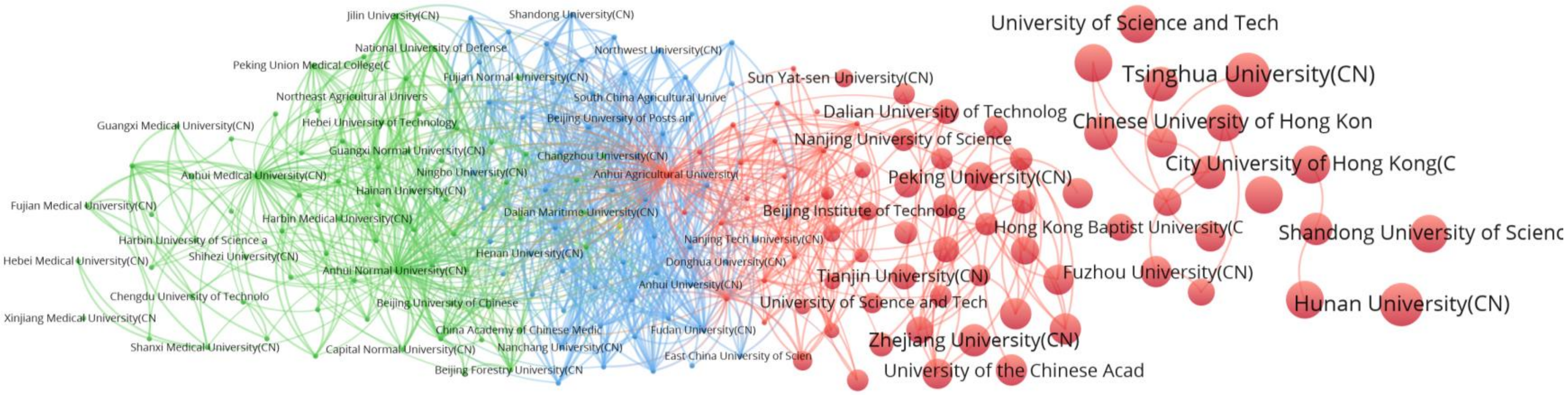

## 5. Comparison among Chinese and American universities

In the LR2020, universities are grouped by nation-states, but it is possible to draw samples of universities across nations or within nations. Classification of universities at lower levels of aggregation can be relevant for the study of regional innovation systems. Whereas universities remain the units of analysis in LR, relevant samples can be drawn from the database on the basis of a specific research question.

For example, if the research question is about comparing universities in the EU, one can study the EU as a single unit internally different from the USA or China. The organization at the level of EU is probably different from the sum of the national perspectives. As noted, LR also contains information for six major fields so that one can cross-tabulate nations with these disciplinary categories, where one could test in principle the conjecture that "China is strong in the basic sciences and the US in the biomedical sciences." We will leave this for a later study, but focus here first on how to compare US and Chinese universities in a single framework.

Both the US and China happen to be represented with approximately 200 universities in LR 2020 (Table 8). Among the 198 American universities, Rockefeller University is an extreme outlier with more than 30% of the papers in the top-10% most-highly cited group. The following analysis is based on 205 Chinese and (198 – 1 (Rockefeller University) =) 197 American universities. The clustering in the network among these (197 + 205 =) 402 universities is visualized in Figure 6. Three clusters and a few isolates are distinguished by the statistical analysis. The isolates are: George Mason University and the University of Toledo in the USA, and the Hangzhou Dianzi University in China. The cross-tabulation in Table 7 is significant at the one-percent level ($\chi^2$ = 93.40; $p$ <.01).

**Table 7**: Descriptive statistics of the comparison among 197 American and 205 Chinese universities

| Row Labels | low | middle | High | Isolates | Grand Total |
|---|---|---|---|---|---|
| China | 116 | 67 | 21 | 1 | 205 |
| USA | 36 | 60 | 99 | 2 | 197 |
| **Grand Total** | **152** | **127** | **120** | | **402** |

**Figure 6**: Grouping of 205 Chinese and 197 US universities in terms of *z*-values; differences among groups are significant at the 1% level; VOSviewer used for the decomposition and clustering.

(The map can be web-started from here.)

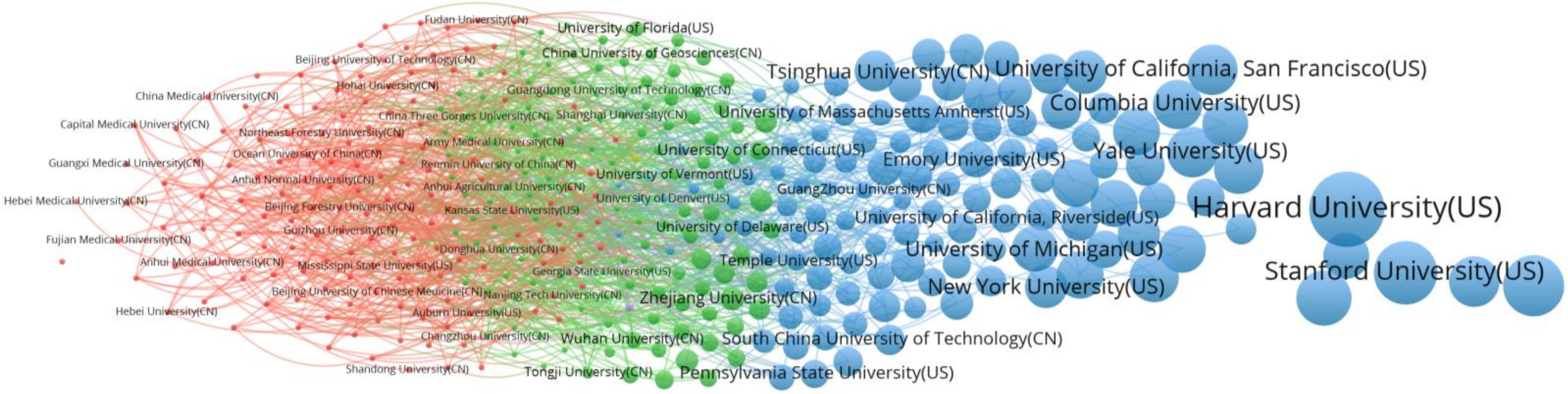

**Table 8:** Highest ranked universities in top-groups in the combined American-Chinese set of 402 universities in LR 2020 with their respective z-scores.

| *rank* | *Top 20 Chinese universities* | *z* | *Top 20 American Universities* | *z* |
|---|---|---|---|---|
| 1 | Tsinghua Univ. | 11.005 | Harvard Univ. | 36.632 |
| 2 | Hunan Univ. | 10.193 | Stanford Univ. | 26.028 |
| 3 | Hong Kong Univ. of Science and Technology | 6.566 | Massachusetts Institute of Technology | 24.504 |
| 4 | Univ. of Science and Technology of China | 6.482 | Univ. of California, Berkeley | 20.319 |
| 5 | City Univ. of Hong Kong | 6.454 | Yale Univ. | 16.576 |
| 6 | Shandong Univ. of Science and Technology | 6.444 | Univ. of California, San Francisco | 16.524 |
| 7 | Hong Kong Polytechnic Univ. | 6.406 | Princeton Univ. | 16.522 |
| 8 | South China Univ. of Technology | 6.049 | Columbia Univ. | 16.348 |
| 9 | Chinese Univ. of Hong Kong | 5.993 | Univ. of California, San Diego | 16.061 |
| 10 | Nankai Univ. | 4.860 | Univ. of Pennsylvania | 15.975 |
| 11 | Univ. of Hong Kong | 4.418 | Cornell Univ. | 14.934 |
| 12 | Shenzhen Univ. | 4.089 | Univ. of Washington, Seattle | 14.620 |
| 13 | Qufu Normal Univ. | 3.724 | Johns Hopkins Univ. | 14.540 |
| 14 | Fuzhou Univ. | 3.487 | Northwestern Univ. | 14.488 |
| 15 | Central China Normal Univ. | 3.090 | Univ. of California, Los Angeles | 14.450 |
| 16 | Univ. of Macau | 3.046 | Univ. of Michigan | 13.951 |
| 17 | Wuhan Univ. of Technology | 2.914 | California Institute of Technology | 13.695 |
| 18 | Southern Univ. of Science and Technology | 2.332 | Univ. of Chicago | 13.561 |
| 19 | GuangZhou Univ. | 2.120 | Duke Univ. | 12.605 |
| 20 | Hong Kong Baptist Univ. | 2.094 | Washington Univ. in St. Louis | 12.123 |

Table 8 shows the top 20 Chinese universities juxtaposed to the 20 American universities with highest *z*-scores at different scales. This table reveals that the highest ranked among the Chinese universities (Tsinghua with $z = 11.005$) does not reach the *z*-level of the lowest among the 20 most-highly ranked American universities in the right column ($z = 12.23$). This may be due in part to historical factors where Chinese authors are not as well integrated into the network of science as others and thus struggle to gain citations, and it may also reflect some quality issues, as well. Figure 7 shows that the distribution of *z*-scores is systematically lower for Chinese universities than for their US counterparts. In other words, using these parameters, China is still far behind the US in terms of the quality of its universities.

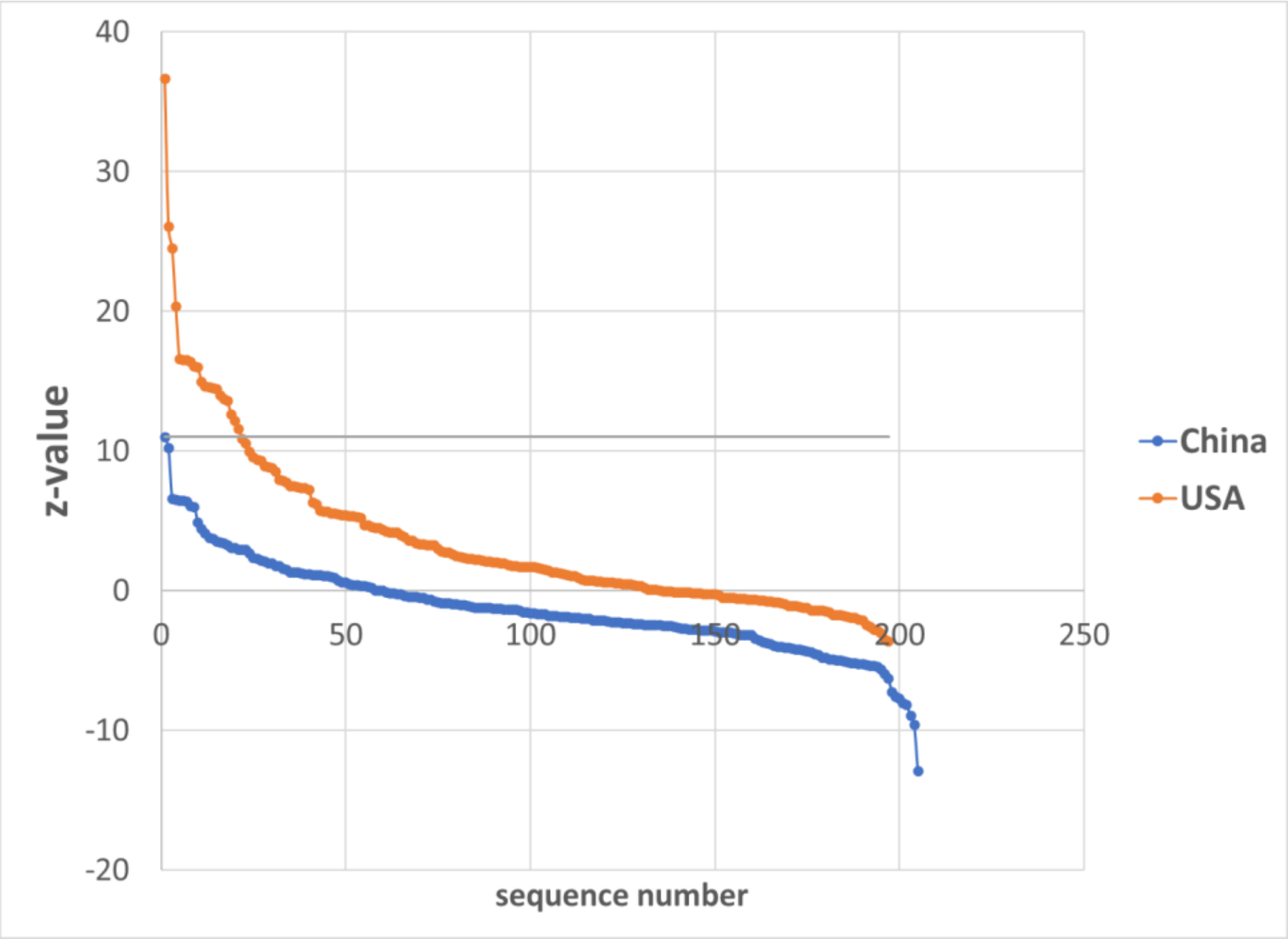

**Figure 7:** Distribution of *z* values in decreasing order for 205 Chinese and 197 American universities.

## 6. Dynamic effects of changes in the model

The methodology for the normalization in terms of different fields of science is continuously improved by CWTS and the database is expanded with new universities. In LR 2016, for example, the number of universities covered was 842 compared to 1076 in this study based on LR2020. The expansion of the database from year to year may have an effect on the rankings because universities may enter the comparison with higher or lower scores on the relevant parameter.

To address the problem of changes in the methodology, LR values are recalculated each year for the historical values of the indicators based on the latest methodology. Thus, we have two time series: one based on the yearly series of LR 2016 to LR2020 and one based on the reconstruction of the data using the method of LR 2020. The two series for Fudan university are graphed in Figure 8.

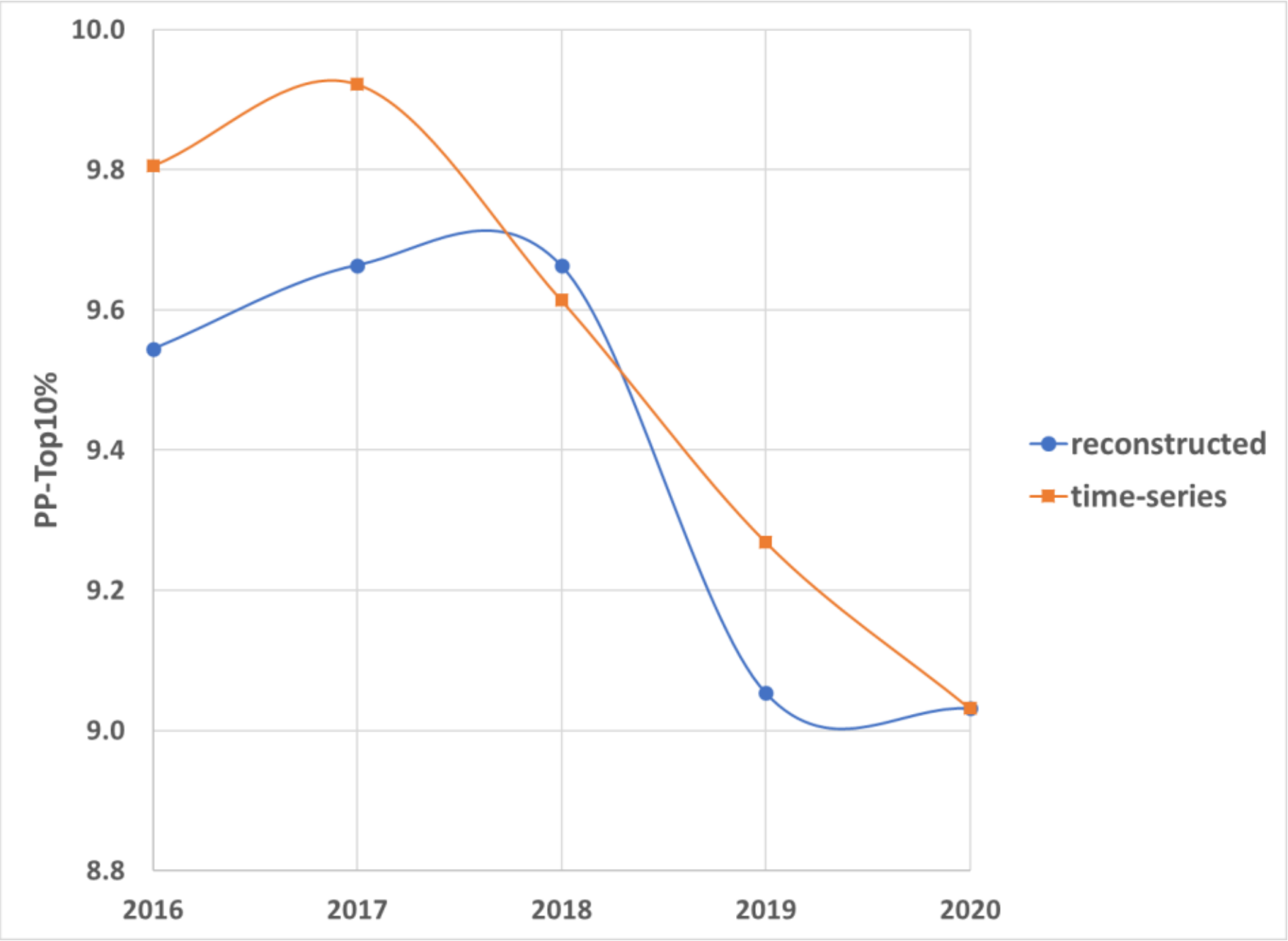

**Figure 8**: The participation of Fudan University in the top-10% class of papers using the Leiden Rankings for subsequent years as a time series *versus* the reconstruction using the 2020-model.

In LR 2016, Fudan University had 9.81% of its papers in the top-10% class (all fields). In 2020, the position of Fudan University has dropped to 9.03 % of its papers in the top-10% class. This is a decline of 0.78% (9.81 – 9.03). Using the 2020 model, however, the value for 2016 is reconstructed as 9.54%, so that we can conclude that the difference in the data is only 0.37% (that is, 9.81 - 9.54%). The remaining decline (9.54 – 9.03 =) 0.51% is an effect of changes in the model. In other words, the model accounts for almost two-thirds of the decline (0.51 / 0.78 = 65.4%) and the citation data themselves for (0.27 / 0.78 =) 34.6%. These relative declines are of the same order of magnitude as the ones shown in Leydesdorff *et al*. (2017) for Carnegie Mellon University in the USA during the period 2012-2016.

## 7. Discussion and conclusions

In summary, both changes in the data and changes in the model can result in differences in the rankings. Whereas scientometric indicators are meant to serve "objectivization" of the discussion about quality, the quality of the indicators themselves is also an issue in the discussion of the results which requires attention. Differences may be due to changes in the data, the models, or the modeling effects on the data.

Our main argument has been that differences among universities can be tested for their statistical significance. This allows for a classification, since universities which are not significantly different can be grouped together, regardless of geographic location. The scientometric groupings, however, should not be reified. The groupings are not stable when we use different methods. Particularly at the margins, the attribution may be sensitive to parameter choices.

The statistics relativize the values of decimals in the rankings. One can operate with a scheme of low/middle/high in policy debates and leave the more fine-grained rankings of individual universities to operational management and local settings. Further analysis can reveal points which merit attention and discussion. Is a decline due to a parameter choice, or is there reason for concern? An alternative view can foreground unseen relationships in the background;

the resulting insights can be made the subject of managerial and political interventions. One can expect that qualitative assessments lag behind the ongoing developments. Cultural expectations are conservative, while all universities are under the pressure to change their position in relationship to one another in a competitive environment. One can zoom in and organize follow-up investigations in these cases. The result may initially be unwelcome, but can also induce asking urgent questions.

At the macro level, our results show that Chinese universities do not yet (?) operate at the same levels of performance as those in the USA. Despite concerns in the U.S. about the ‘competition’ from Chinese universities, the latter do not rank in the same elite categories as American universities at this time. As in other national systems, we found three or four groups of universities at national levels. In the case of China, the top-list includes five universities with an address in in Hong Kong and perhaps partly because of the historical British tradition operating in the background.

**Acknowledgments:** Lin Zhang acknowledges support of the National Natural Science Foundation of China (Grant No. 71974150).

## Annex 1: Top group of 32 Chinese universities

| ***University*** | ***Z*** | ***overall rank*** | ***within-group rank*** |
|---|---|---|---|
| Tsinghua University | 11.005 | 1 | 1 |
| Hunan University | 10.193 | 2 | 2 |
| Hong Kong University of Science and Technology | 6.566 | 3 | 3 |
| University of Science and Technology of China | 6.482 | 4 | 4 |
| City University of Hong Kong | 6.454 | 5 | 5 |
| Shandong University of Science and Technology | 6.444 | 6 | 6 |
| Hong Kong Polytechnic University | 6.406 | 7 | 7 |
| South China University of Technology | 6.049 | 8 | 8 |
| Chinese University of Hong Kong | 5.993 | 9 | 9 |
| Nankai University | 4.860 | 10 | 10 |
| University of Hong Kong | 4.418 | 11 | 11 |
| Shenzhen University | 4.089 | 12 | 12 |
| Qufu Normal University | 3.724 | 14 | 13 |
| Fuzhou University | 3.487 | 15 | 14 |
| Beihang University | 3.432 | 16 | 15 |
| University of the Chinese Academy of Sciences | 3.354 | 17 | 16 |
| Central China Normal University | 3.090 | 19 | 17 |
| University of Macau | 3.046 | 20 | 18 |
| Northwestern Polytechnical University | 2.915 | 21 | 19 |
| Wuhan University of Technology | 2.914 | 22 | 20 |
| Southern University of Science and Technology | 2.332 | 25 | 21 |
| Beijing University of Chemical Technology | 2.273 | 26 | 22 |
| GuangZhou University | 2.120 | 27 | 23 |
| Hong Kong Baptist University | 2.094 | 28 | 24 |
| Shandong Normal University | 1.478 | 34 | 25 |
| Jiangsu Normal University | 1.313 | 36 | 26 |
| Zhejiang Normal University | 1.238 | 38 | 27 |
| University of Jinan | 1.195 | 39 | 28 |
| Nanjing University of Information Science and Technology | 1.093 | 42 | 29 |
| Hangzhou Dianzi University | 0.449 | 51 | 30 |
| Heilongjiang University | -0.158 | 62 | 31 |
| Qingdao Agricultural University | -0.422 | 67 | 32 |

## Annex 2: Middle group of 69 Chinese universities

| *University* | *z* | *overall rank* | *within-group* |
|---|---|---|---|
| Zhejiang University | 3.800 | 13 | 1 |
| Harbin Institute of Technology | 3.226 | 18 | 2 |
| Huazhong University of Science and Technology | 2.907 | 23 | 3 |
| Peking University | 2.689 | 24 | 4 |
| Nanjing University | 1.973 | 29 | 5 |
| Xiamen University | 1.954 | 30 | 6 |
| Wuhan University | 1.765 | 31 | 7 |
| Tianjin University | 1.738 | 32 | 8 |
| Dalian University of Technology | 1.553 | 33 | 9 |
| East China Normal University | 1.333 | 35 | 10 |
| Soochow University | 1.310 | 37 | 11 |
| University of Electronic Science and Technology of China | 1.154 | 40 | 12 |
| Nanjing University of Science and Technology | 1.140 | 41 | 13 |
| Southwest Jiaotong University | 1.088 | 43 | 14 |
| China University of Geosciences | 1.035 | 44 | 15 |
| Huazhong Agricultural University | 1.030 | 45 | 16 |
| Tongji University | 0.954 | 46 | 17 |
| University of Science and Technology Beijing | 0.932 | 47 | 18 |
| China Agricultural University | 0.704 | 48 | 19 |
| Beijing Institute of Technology | 0.587 | 49 | 20 |
| Sun Yat-sen University | 0.572 | 50 | 21 |
| North China Electric Power University | 0.414 | 52 | 22 |
| China University of Petroleum Beijing | 0.389 | 53 | 23 |
| Guangdong University of Technology | 0.332 | 54 | 24 |
| Shanghai University | 0.332 | 55 | 25 |
| Fourth Military Medical University | 0.257 | 56 | 26 |
| Beijing Normal University | 0.178 | 57 | 27 |
| Chongqing University | 0.000 | 58 | 28 |
| Hefei University of Technology | 0.000 | 59 | 29 |
| Xi'an Jiaotong University | 0.000 | 60 | 30 |
| Nanjing University of Posts and Telecommunications | -0.110 | 61 | 31 |
| Nanjing Agricultural University | -0.171 | 63 | 32 |
| China Pharmaceutical University | -0.259 | 64 | 33 |
| Central South University | -0.280 | 65 | 34 |
| China University of Petroleum East China | -0.407 | 66 | 35 |
| Beijing Jiaotong University | -0.448 | 68 | 36 |
| Xidian University | -0.529 | 71 | 37 |
| Dalian Medical University | -0.643 | 72 | 38 |
| Chongqing University of Posts and Telecommunications | -0.656 | 73 | 39 |
| Huaqiao University | -0.795 | 74 | 40 |
| Anhui Agricultural University | -0.862 | 75 | 41 |
| Wuhan University of Science and Technology | -0.899 | 76 | 42 |
| Nanjing Tech University | -0.919 | 77 | 43 |
| Donghua University | -0.923 | 78 | 44 |
| East China University of Science and Technology | -0.943 | 79 | 45 |
| Army Medical University | -0.952 | 80 | 46 |
| Anhui University | -1.004 | 81 | 47 |
| Tianjin University of Technology | -1.047 | 82 | 48 |
| Renmin University of China | -1.064 | 83 | 49 |
| Nanjing Normal University | -1.189 | 84 | 50 |
| Southeast University | -1.213 | 85 | 51 |
| Fujian Agriculture and Forestry University | -1.218 | 86 | 52 |
| Northwest Agriculture and Forestry University | -1.226 | 87 | 53 |
| University of Shanghai for Science and Technology | -1.232 | 88 | 54 |
| North China University of Science and Technology | -1.249 | 89 | 55 |
| Wenzhou University | -1.306 | 90 | 56 |
| Southwest Petroleum University | -1.311 | 91 | 57 |
| Shaanxi University of Science and Technology | -1.314 | 92 | 58 |
| China Three Gorges University | -1.333 | 93 | 59 |
| University of South China | -1.345 | 94 | 60 |
| Shanghai Normal University | -1.369 | 95 | 61 |
| Dalian Maritime University | -1.380 | 96 | 62 |
| Yangzhou University | -1.392 | 97 | 63 |
| Changzhou University | -1.544 | 98 | 64 |
| Nanjing Forestry University | -1.581 | 99 | 65 |
| Guangzhou University of Chinese Medicine | -1.685 | 104 | 66 |
| Fujian Normal University | -1.781 | 105 | 67 |
| China Academy of Chinese Medical Sciences | -1.840 | 107 | 68 |
| Shanghai Jiao Tong University | -1.876 | 109 | 69 |

## Annex 3: Bottom group of 102 Chinese universities

| *University* | *z* | *rank* | *within-group* |
|---|---|---|---|
| Hubei University | -0.498 | 70 | 1 |
| Northwest University | -1.609 | 100 | 2 |
| South China Agricultural University | -1.631 | 101 | 3 |
| Zhejiang University of Technology | -1.676 | 102 | 4 |
| China University of Mining and Technology | -1.678 | 103 | 5 |
| Nanjing University of Aeronautics and Astronautics | -1.816 | 106 | 6 |
| Second Military Medical University | -1.844 | 108 | 7 |
| Zhejiang Sci-Tech University | -1.886 | 110 | 8 |
| Taiyuan University of Technology | -1.934 | 111 | 9 |
| Shanghai University of Traditional Chinese Medicine | -1.950 | 112 | 10 |
| Qingdao University of Science and Technology | -1.961 | 113 | 11 |
| Nanjing University of Chinese Medicine | -1.981 | 114 | 12 |
| Harbin Engineering University | -2.033 | 115 | 13 |
| Lanzhou University | -2.110 | 117 | 14 |
| Northeast Normal University | -2.128 | 118 | 15 |
| China Jiliang University | -2.130 | 119 | 16 |
| Jiangnan University | -2.163 | 120 | 17 |
| Jinan University | -2.227 | 121 | 18 |
| Hangzhou Normal University | -2.245 | 122 | 19 |
| Southern Medical University | -2.263 | 123 | 20 |
| Hunan Normal University | -2.282 | 124 | 21 |
| Henan University | -2.328 | 125 | 22 |
| Beijing University of Chinese Medicine | -2.346 | 126 | 23 |
| Tianjin Medical University | -2.348 | 127 | 24 |
| Shantou University | -2.383 | 128 | 25 |
| Xi'an University of Architecture and Technology | -2.384 | 129 | 26 |
| Hainan University | -2.417 | 130 | 27 |
| Hohai University | -2.453 | 131 | 28 |
| Beijing University of Posts and Telecommunications | -2.458 | 132 | 29 |
| Qingdao University | -2.465 | 133 | 30 |
| Jiangsu University | -2.467 | 134 | 31 |
| Jiangxi Normal University | -2.486 | 135 | 32 |
| Shaanxi Normal University | -2.505 | 136 | 33 |
| Shandong University of Technology | -2.506 | 137 | 34 |
| Nanchang University | -2.528 | 138 | 35 |
| Henan Agricultural University | -2.584 | 139 | 36 |
| Guizhou University | -2.675 | 140 | 37 |
| Guangxi Normal University | -2.709 | 141 | 38 |
| Northeast Forestry University | -2.751 | 142 | 39 |
| Northeastern University | -2.761 | 143 | 40 |
| Henan Normal University | -2.775 | 144 | 41 |
| Shandong Agricultural University | -2.783 | 145 | 42 |
| Anhui Normal University | -2.834 | 146 | 43 |
| Lanzhou University of Technology | -2.835 | 147 | 44 |
| Ningbo University | -2.838 | 148 | 45 |
| Southwest University | -2.855 | 149 | 46 |
| Yunnan University | -2.940 | 150 | 47 |
| Shanghai Ocean University | -2.954 | 151 | 48 |
| Northwest Normal University | -2.962 | 152 | 49 |
| Inner Mongolia University | -2.968 | 153 | 50 |
| Beijing University of Technology | -2.969 | 154 | 51 |
| Fudan University | -3.065 | 155 | 52 |
| Yanshan University | -3.100 | 156 | 53 |
| Tianjin University of Science and Technology | -3.150 | 157 | 54 |
| Beijing Forestry University | -3.151 | 158 | 55 |
| People's Liberation Army University of Science and Technology | -3.152 | 159 | 56 |
| Capital Normal University | -3.188 | 160 | 57 |
| Southwest University of Science and Technology | -3.433 | 161 | 58 |
| Hebei University of Technology | -3.558 | 162 | 59 |
| Shenyang Pharmaceutical University | -3.713 | 163 | 60 |
| Xinjiang University | -3.741 | 164 | 61 |
| Kunming University of Science and Technology | -3.832 | 165 | 62 |
| Nantong University | -3.935 | 166 | 63 |
| Xiangtan University | -4.012 | 167 | 64 |
| Northeast Agricultural University | -4.025 | 168 | 65 |
| Wenzhou Medical University | -4.062 | 169 | 66 |
| Chongqing Medical University | -4.106 | 170 | 67 |
| Guangxi University | -4.119 | 171 | 68 |
| Xi'An University of Technology | -4.203 | 172 | 69 |
| Zhengzhou University | -4.244 | 173 | 70 |
| Ocean University of China | -4.246 | 174 | 71 |
| Henan Polytechnic University | -4.348 | 175 | 72 |
| Shihezi University | -4.400 | 176 | 73 |
| Kunming Medical University | -4.506 | 177 | 74 |
| Xuzhou Medical College | -4.604 | 178 | 75 |
| Tianjin Polytechnic University | -4.775 | 179 | 76 |
| South China Normal University | -4.827 | 180 | 77 |
| Nanjing Medical University | -4.941 | 181 | 78 |
| Chengdu University of Technology | -4.958 | 182 | 79 |
| Xinxiang Medical University | -4.962 | 183 | 80 |
| Harbin University of Science and Technology | -5.001 | 184 | 81 |
| North University of China | -5.068 | 185 | 82 |
| Yangtze University | -5.093 | 186 | 83 |
| Guangzhou Medical University | -5.200 | 187 | 84 |
| Sichuan Agricultural University | -5.212 | 188 | 85 |
| Sichuan University | -5.247 | 189 | 86 |
| National University of Defense Technology | -5.257 | 190 | 87 |
| Harbin Medical University | -5.331 | 191 | 88 |
| Tianjin Normal University | -5.358 | 192 | 89 |
| Shanxi Medical University | -5.378 | 193 | 90 |
| Henan University of Science and Technology | -5.416 | 194 | 91 |
| Shandong University | -5.619 | 195 | 92 |
| Anhui Medical University | -5.951 | 196 | 93 |
| Hebei University | -6.307 | 197 | 94 |
| Guangxi Medical University | -7.267 | 198 | 95 |
| Peking Union Medical College | -7.621 | 199 | 96 |
| Xinjiang Medical University | -7.703 | 200 | 97 |
| China Medical University | -8.064 | 201 | 98 |
| Fujian Medical University | -8.169 | 202 | 99 |
| Hebei Medical University | -8.966 | 203 | 100 |
| Jilin University | -9.607 | 204 | 101 |
| Capital Medical University | -12.944 | 205 | 102 |

**Annex 4: 75 top universities with overlapping confidence values**

| *University* | *Z* | *rank* | *within group* |
|---|---|---|---|
| Tsinghua University | 11.005 | 1 | 1 |
| Hunan University | 10.193 | 2 | 2 |
| Hong Kong University of Science and Technology | 6.566 | 3 | 3 |
| University of Science and Technology of China | 6.482 | 4 | 4 |
| City University of Hong Kong | 6.454 | 5 | 5 |
| Shandong University of Science and Technology | 6.444 | 6 | 6 |
| Hong Kong Polytechnic University | 6.406 | 7 | 7 |
| South China University of Technology | 6.049 | 8 | 8 |
| Chinese University of Hong Kong | 5.993 | 9 | 9 |
| Nankai University | 4.860 | 10 | 10 |
| University of Hong Kong | 4.418 | 11 | 11 |
| Shenzhen University | 4.089 | 12 | 12 |
| Zhejiang University | 3.800 | 13 | 13 |
| Qufu Normal University | 3.724 | 14 | 14 |
| Fuzhou University | 3.487 | 15 | 15 |
| Beihang University | 3.432 | 16 | 16 |
| University of the Chinese Academy of Sciences | 3.354 | 17 | 17 |
| Harbin Institute of Technology | 3.226 | 18 | 18 |
| Central China Normal University | 3.090 | 19 | 19 |
| University of Macau | 3.046 | 20 | 20 |
| Northwestern Polytechnical University | 2.915 | 21 | 21 |
| Wuhan University of Technology | 2.914 | 22 | 22 |
| Huazhong University of Science and Technology | 2.907 | 23 | 23 |
| Peking University | 2.689 | 24 | 24 |
| Southern University of Science and Technology | 2.332 | 25 | 25 |
| Beijing University of Chemical Technology | 2.273 | 26 | 26 |
| GuangZhou University | 2.120 | 27 | 27 |
| Hong Kong Baptist University | 2.094 | 28 | 28 |
| Nanjing University | 1.973 | 29 | 29 |
| Xiamen University | 1.954 | 30 | 30 |
| Wuhan University | 1.765 | 31 | 31 |
| Tianjin University | 1.738 | 32 | 32 |
| Dalian University of Technology | 1.553 | 33 | 33 |
| Shandong Normal University | 1.478 | 34 | 34 |
| East China Normal University | 1.333 | 35 | 35 |
| Jiangsu Normal University | 1.313 | 36 | 36 |
| Soochow University | 1.310 | 37 | 37 |
| Zhejiang Normal University | 1.238 | 38 | 38 |
| University of Jinan | 1.195 | 39 | 39 |
| University of Electronic Science and Technology of China | 1.154 | 40 | 40 |
| Nanjing University of Science and Technology | 1.140 | 41 | 41 |
| Nanjing University of Information Science and Technology | 1.093 | 42 | 42 |
| Southwest Jiaotong University | 1.088 | 43 | 43 |
| China University of Geosciences | 1.035 | 44 | 44 |
| Huazhong Agricultural University | 1.030 | 45 | 45 |
| Tongji University | 0.954 | 46 | 46 |
| University of Science and Technology Beijing | 0.932 | 47 | 47 |
| China Agricultural University | 0.704 | 48 | 48 |
| Beijing Institute of Technology | 0.587 | 49 | 49 |
| Sun Yat-sen University | 0.572 | 50 | 50 |
| Hangzhou Dianzi University | 0.449 | 51 | 51 |
| North China Electric Power University | 0.414 | 52 | 52 |
| China University of Petroleum Beijing | 0.389 | 53 | 53 |
| Guangdong University of Technology | 0.332 | 54 | 54 |
| Shanghai University | 0.332 | 55 | 55 |
| Fourth Military Medical University | 0.257 | 56 | 56 |
| Beijing Normal University | 0.178 | 57 | 57 |
| Chongqing University | 0.000 | 58 | 58 |
| Hefei University of Technology | 0.000 | 59 | 59 |
| Xi'an Jiaotong University | 0.000 | 60 | 60 |
| Nanjing University of Posts and Telecommunications | -0.110 | 61 | 61 |
| Heilongjiang University | -0.158 | 62 | 62 |
| Nanjing Agricultural University | -0.171 | 63 | 63 |
| China Pharmaceutical University | -0.259 | 64 | 64 |
| Central South University | -0.280 | 65 | 65 |
| China University of Petroleum East China | -0.407 | 66 | 66 |
| Qingdao Agricultural University | -0.422 | 67 | 67 |
| Beijing Jiaotong University | -0.448 | 68 | 68 |
| Chang’an University | -0.465 | 69 | 69 |
| Hubei University | -0.498 | 70 | 70 |
| Dalian Medical University | -0.643 | 72 | 71 |
| Chongqing University of Posts and Telecommunications | -0.656 | 73 | 72 |
| Huaqiao University | -0.795 | 74 | 73 |
| Anhui Agricultural University | -0.862 | 75 | 74 |
| Army Medical University | -0.952 | 80 | 75 |